\documentclass[fleqn,usenatbib]{rasti}

\usepackage{newtxtext,newtxmath}

\usepackage[T1]{fontenc}

\DeclareRobustCommand{\VAN}[3]{#2}
\let\VANthebibliography\thebibliography
\def\thebibliography{\DeclareRobustCommand{\VAN}[3]{##3}\VANthebibliography}

\usepackage{graphicx}	
\usepackage{amsmath}	

\newcommand{\studentt}[2]{t_\nu \left( #1, #2 \right)}
\newcommand{\depvar}{y_i}
\newcommand{\indepvars}{\boldsymbol{x}_i}
\newcommand{\obsdep}{\hat{y}_i}
\newcommand{\obsindep}{\hat{\boldsymbol{x}}_i}
\newcommand{\indepcov}{\Sigma_{x, i}}
\newcommand{\deperr}{\sigma_{y, i}}
\newcommand{\intscttr}{\sigma_{\text{int}}}
\newcommand{\intercept}{\alpha}
\newcommand{\covariate}{\beta}

\title[Robust regression in astronomy]{An approach to robust Bayesian regression in astronomy}

\author[W. Martin \& D. Mortlock]{
William Martin$^{1}$\thanks{E-mail: \href{mailto:w.martin19@imperial.ac.uk}{w.martin19@imperial.ac.uk}}
and Daniel Mortlock$^{1,2}$
\\
$^{1}$Department of Physics, Imperial College London, Blackett Laboratory, Prince Consort Road, London SW7 2AZ, UK\\
$^{2}$Department of Mathematics, Imperial College London, London, SW7 2AZ, UK
}

\date{Accepted 31/07/2025. Received 29/07/2025; in original form 21/11/2024}

\pubyear{\the\year{}}

\begin{document}
\label{firstpage}
\pagerange{\pageref{firstpage}--\pageref{lastpage}}
\maketitle

\begin{abstract}
Model mis-specification (e.g.\ the presence of outliers) is commonly encountered
in astronomical analyses, often requiring the use of ad hoc algorithms  which
are sensitive to arbitrary thresholds (e.g.\ sigma-clipping). For any given
dataset, the optimal approach will be to develop a bespoke statistical model of
the data generation and measurement processes, but these come with a development
cost; there is hence utility in having generic modelling approaches that are
both principled and robust to model mis-specification. Here we develop and
implement a generic Bayesian approach to linear regression, based on Student's
$t$-distributions, that is robust to outliers and mis-specification of the noise
model. Our method is validated using simulated datasets with various degrees of
model mis-specification; the derived constraints are shown to be systematically
less biased than those from a similar model using normal distributions. We
demonstrate that, for a dataset without outliers, a worst-case inference using
$t$-distributions would give unbiased results with $\lesssim\!\!10$ per cent
increase in the reported parameter uncertainties. We also compare with existing
analyses of real-world datasets, finding qualitatively different results where
normal distributions have been used and agreement where more robust methods have
been applied. A Python implementation of this model, $t$-cup, is made available
for others to use.
\end{abstract}




\section{Introduction}
\label{sec:intro}









Linear regression is a common problem in astronomy, arising in fields as diverse
as galaxy formation and evolution (e.g.\ the super-massive black hole (SMBH)
mass -- stellar velocity dispersion correlation, \citealt{Ferrarese:2000,
Gebhardt:2000}), stellar physics (e.g.\ the Leavitt law linking the luminosity
and pulsation period for Cepheid variable stars, \citealt{Leavitt:1912}), and
cosmology (e.g.\ the original formulation of Hubble's law,
\citealt{Hubble:1929}). For this reason, there are a plethora of techniques used
by astronomers for linear regression when both the dependent and independent
variables are measured with error (e.g., \citealt{Press:1992, Akritas:1996,
Tremaine:2002, Kelly:2007} -- see \citealt{Andreon:2013} and
\citealt{Andreon:2015} for reviews including further examples).

\citet{Kelly:2007} illustrates that common ad-hoc estimators such as
\textsc{fitexy} \citep{Press:1992, Tremaine:2002} and \textsc{bces}
\citep{Akritas:1996} suffer from biases and can underestimate intrinsic scatter.
Similar algorithms exist for removing outliers from data (e.g.\ sigma clipping)
have been used previously but these can lead to controversy (e.g.\ the
reanalysis of the data-set from \citealt{Riess:2011} by
\citealt{Efstathiou:2014}). A more principled approach is to model the entire
measurement process.

Bayesian hierarchical models (BHMs) are a natural way to model such datasets --
these allow astronomers to account for, e.g., measurement errors, selection
effects, interlinked parameters, censored data, and many other effects common to
astronomical problems. BHMs have seen increasing use in astrophysics over the
past few decades, from the distance-redshift relation in cosmology
\citep[e.g.][]{Feeney:2018, Avelino:2019} and photometric redshift estimation
\citep[e.g.][]{Leistedt:2016} to exoplanet characterisation
\citep[e.g.][]{Sestovic:2018} and population-level inference
\citep[e.g.][]{Kelly:2009} -- see \citet{Feigelson:2021} for a recent review
including further examples of BHMs.

\citet{Kelly:2007} presented a general BHM for linear regression with
measurement errors and censored data, demonstrating  several advantages over the
other methods considered: no bootstrapping was required to obtain uncertainties
on parameters; the Bayesian approach was easily extensible to truncated or
censored data; and other methods would sometimes severely underestimate
intrinsic scatter in the data. This formulation of Bayesian regression,
sometimes known as \textsc{linmix\_err}, is now commonly used in astronomy
\citep[e.g.][]{McConnell:2013, Bentz:2013, Andrews:2013}. This approach has been
extended and refined by others \citep[e.g.][]{Mantz:2016, Sereno:2016,
Bartlett:2023, Jing:2024}. These
models assume parameters are normally distributed throughout; however,
scientific uncertainties are often empirically not normally distributed, leading
to more frequent outliers \citep{Bailey:2017}. Inference that relies on normal
distributions can be unduly affected by outliers (see Section~\ref{sec:results}
for an exploration of this effect).

The problem of outliers within datasets can be thought of as model
mis-specification: these objects do not fit the distributions used to model
them. At very least, results should be checked by performing analyses using
different models for the noise distribution \citep[e.g.][]{Andreon:2015,
McElreath:2020}. However, it would be preferable to use robust inference
methods, which are designed to work irrespective of the actual generative
distribution \citep{Berger:1994}. One approach is to use distributions that are
leptokurtic (i.e.\ have heavier tails than a normal distribution) for inference,
with suggestions including Student's $t$-distributions \citep{Andrews:1974},
Gaussian mixture models \citep{Box:1968, Aitkin:1980} and log-regularly varying
distributions \citep[LRVD;][]{Gagnon:2020, Hamura:2022}, all of which have been
successful in practice \citep[e.g.][]{Berger:1994, Sivia:2006, Gelman:2013,
Gagnon:2023, Hamura:2024}.  Student's $t$-distributions, parameterized by the
shape parameter $\nu$, have seen use in bespoke astronomical and cosmological
inference, both by fixing $\nu$ \citep[e.g.][]{Andreon:2008, Jontof-Hutter:2016,
Andreon:2020} and treating $\nu$ as a free parameter \citep[e.g.][]{Park:2017,
Feeney:2018}. The advantage of the latter approach is that we can marginalise
over $\nu$ so that we only consider models supported by the data
\citep{Gelman:2013}.  While there are code examples in which Student's
$t$-distributions are used for the scatter term in the dependent variable of
regression models (e.g.\ Appendix~C.2 of \citealt{Gelman:2013}, Section~9.1.3 of
\citealt{Andreon:2015}, or Example~7.35 of \citealt{McElreath:2020}), there is
not currently a general method for robust Bayesian regression in astronomy which
incorporates uncertainties in all quantities.

We propose a development of a generic approach for robust astronomical data
analysis. From the review of previous methods, we can identify properties that
we would like to see in our regression model:

\begin{itemize}
	\item A BHM -- we favour a hierarchical approach because it naturally
	encodes the hierarchical structure of astronomical regression problems
	(i.e.\ objects are drawn from a high-level population; the objects
	intrinsically obey some relationship; the objects are measured with error
	and only the measured values are known). We adopt a Bayesian approach both
	for practical reasons -- the resultant posterior distributions provide full
	uncertainty quantification -- and due to their logical consistency
	\citep[e.g.][]{Cox:1946, vanHorn:2003, Knuth:2010}.

	\item A robust model -- we desire a model that is robust to both outliers
	(i.e.\ data points that, whether intrinsically or by virtue of measurement
	errors, do not lie sufficiently close to our regression relation) and model
	mis-specification (i.e.\ where the underlying distribution of data does not
	match up with the distribution assumed for modelling).

	\item A general method -- we seek a model that does not require case-by-case
	optimisation for application to different regression problems (e.g.\ no
	manual outlier identification and removal, no need to rescale prior
	distributions for different problems, etc.). While a bespoke model tailored
	to a specific problem will perform better than a general method, there is
	still value in making a generic approach available for analysis.
\end{itemize}

We implement these ideas in this paper, beginning with a discussion of our model
in Section~\ref{sec:formalism}. In Section~\ref{sec:methods}, we outline the
methods that we use to validate our model; the results of these validation
checks are presented in Section~\ref{sec:results}. In
Section~\ref{sec:real-world}, we compare the performance of our model on
real-world datasets with the models outlined in \citet{Kelly:2007} and
\citet{Park:2017}, before summarizing our conclusions in
Section~\ref{sec:conclusion}.

\section{Formalism}
\label{sec:formalism}

Here we establish notation and set out our model. Our dataset has $N$
astronomical objects, each with a $K$-dimensional vector of associated
independent quantities $\{\boldsymbol{x}_i\}$ and a dependent quantity
$\{y_i\}$. For example, if we were estimating SMBH mass using measurements of
luminosity and line width, we would have $K = 2$ dimensional vectors of
independent quantities $\{\boldsymbol{x}_i\} = \{(L_i, \Delta V_i)^T\}$, and the
dependent quantity $\{y_i\}$ would correspond to the black hole mass.

We assume that the independent quantities $\{\boldsymbol{x}_i\}$ and dependent
quantities $\{y_i\}$ are related by a regression relation
\begin{align}
    y_i =&\; f(\boldsymbol{x}_i; \boldsymbol{\theta}_f) + \delta_i, \\
    \delta_i \sim&\;
       \mathcal{P}_{\text{int}} \left( \boldsymbol{\theta}_{\text{int}} \right),
\end{align}
where $f(\boldsymbol{x}_i; \boldsymbol{\theta}_f)$ is a function relating
$\{\boldsymbol{x}_i\}$ to $\{y_i\}$, with parameters $\boldsymbol{\theta}_f$,
and $\mathcal{P}_{\text{int}}$ is an unknown probability distribution with
parameters $\boldsymbol{\theta}_{\text{int}}$.

These objects are then observed, resulting in the measured data
\begin{align}
    \hat{\boldsymbol{x}}_i =&\;
        \boldsymbol{x}_i + \boldsymbol{\epsilon}_{x,i}, \\
    \hat{y}_i =&\; y_i + \epsilon_{y,i}, \\
    \boldsymbol{\epsilon}_{x,i} \sim&\;
        \mathcal{P}_{\text{obs}} \left(
            \boldsymbol{\theta}_{\text{obs}}
        \right), \\
    \epsilon_{y,i} \sim&\;
        \mathcal{P}_{\text{obs}} \left(
            \boldsymbol{\theta}_{\text{obs}}
        \right),
\end{align}
where $\mathcal{P}_{\text{obs}}$ is an unknown probability distribution with
parameters $\boldsymbol{\theta}_{\text{obs}}$.

When building a model to infer parameters of interest for our model, we must
choose probability distributions to represent the unknown
$\mathcal{P}_{\text{int}}$ and $\mathcal{P}_{\text{obs}}$ -- we label the chosen
distributions as $\tilde{\mathcal{P}}_{\text{int}}$ and
$\tilde{\mathcal{P}}_{\text{obs}}$.

In a Bayesian framework, we can extend this model to deal with, e.g., censored
data or selection effects, but this is beyond the scope of the current work --
see \citet{Kelly:2007} for an overview of an approach that would incorporate
these effects.

\subsection{Building a robust model}
\label{sec:formalism.robust}

In the setup outlined above, the form of the ``true'' distributions
$\mathcal{P}_{\text{int}}$ and $\mathcal{P}_{\text{obs}}$ are unknown; we can
only choose the forms of $\tilde{\mathcal{P}}_{\text{int}}$ and
$\tilde{\mathcal{P}}_{\text{obs}}$. A common assumption in analysis is to use
normal distributions to model both of these.  However, in the case of model
misspecification (where, e.g., we assume $\tilde{\mathcal{P}}_{\text{int}}$ is a
normal distribution but the true $\mathcal{P}_{\text{int}}$ is a different
distribution), the results obtained under this assumption can be biased. Making
the assumption that $\tilde{\mathcal{P}}_{\text{int}}$ follows a different
distribution can give results that are robust to this model misspecification.
This means that, even though we do not believe that the distribution we choose
to model $\tilde{\mathcal{P}}_{\text{int}}$ is the same as the underlying,
unknown $\mathcal{P}_{\text{int}}$, we can have greater confidence in the
inferences about the regression model and the properties of individual objects.

In choosing the form of $\tilde{\mathcal{P}}_{\text{int}}$, some consideration
must be given to the type of outliers which are expected. In the case of extreme
outliers, it is possible to use distributions with very heavy tails such as LRVD
to provide whole robustness \citep{Gagnon:2020, Hamura:2022}. However, the focus
here is on the less extreme situation of more numerous but moderate outliers
that cannot be identified as easily but will nonetheless affect the inference.
For this reason, we use Student's $t$-distributions, which have sufficient
flexibility to represent any noise model which is unimodal and approximately
symmetric. The model presented will not alleviate all possible issues of model
misspecification, such as a strongly misspecified regression relation
$f(\mathbf{x}_i; \mathbf{\theta}_f)$, but, as we shall show in
Section~\ref{sec:results}, provides markedly improved performance over a similar
model with normal distributions, while maintaining constraining power.

\subsection{Sampling distribution}
\label{sec:formalism.sampling}

For robust inference, we seek a sampling distribution that can have heavier
tails than a normal distribution, but that can reduce to a normal distribution
when the underlying dataset is normally distributed. We further seek an
identifiable model, and a differentiable distribution which can be fit using
Hamiltonian Monte Carlo (HMC). These constraints lead us naturally to Student's
$t$-distributions, which fulfil all of these criteria.

Student's $t$-distributions are encountered when estimating the mean of a normal
distribution with unknown variance from a limited number of samples. The number
of samples is a parameter of the distribution: for $n$ samples, the
corresponding Student's $t$-distribution will have $\nu \equiv n - 1$
``degrees-of-freedom''. This value of $\nu$ parameterizes how heavy-tailed the
distribution is. While the interpretation of $\nu$ as ``degrees-of-freedom''
only makes sense for $\nu \in \mathbb Z^+$, the distribution is normalizable for
any positive, real $\nu$; for this reason, we shall refer to $\nu$ as the shape
parameter.

\begin{figure}
	\includegraphics{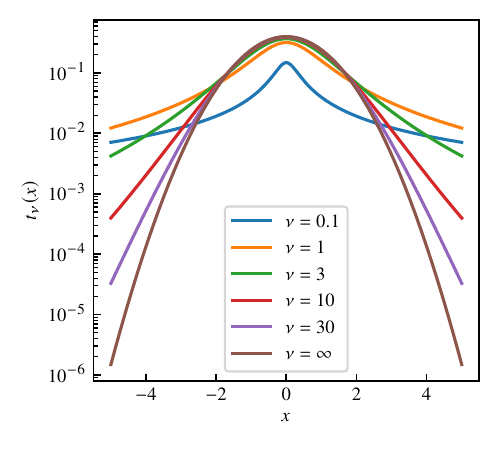}
    \caption{The $t$-distribution probability density function for different
    values of shape parameter, $\nu$.}
    \label{fig:model.t}
\end{figure}

The Student's $t$-distribution, with location $\mu$ and scale $\sigma$, has
the probability density function
\begin{equation}
    t_{\nu} \left(x; {\mu}, {\sigma^2}\right)
        =
    \frac{1}{\sqrt{\pi \nu} \sigma}
    \frac{
        \Gamma \left(\frac{\nu + 1}2\right)
    }{
        \Gamma \left(\frac{\nu}2\right)
    }
    \left(
        1 + \frac{1}{\nu} \frac{\left(x - \mu\right)^2}{\sigma^2}
    \right)^{
        -\frac{\nu + 1}{2}
    }.
\end{equation}
This is shown for a range of $\nu$ in Figure~\ref{fig:model.t}: $\nu
= 1$ gives a Cauchy distribution; and $\nu \rightarrow \infty$ tends to a normal
distribution. The distribution has mean
\begin{equation}
    \mathbb{E}(x)
        =
    \begin{cases}
        \mu & \nu > 1, \\
        \textrm{undefined} & \textrm{otherwise}
    \end{cases}
\end{equation}
and variance
\begin{equation}
    \mathrm{Var}(x)
        =
    \begin{cases}
        \frac{\nu}{\nu - 2} \sigma^2 & \nu > 2, \\
        \infty & 1 < \nu \leq 2, \\
        \textrm{undefined} & \textrm{otherwise.}
    \end{cases}
\end{equation}

In this paper, we have found it useful to define two quantities for comparison
with normal distributions. The first, $\sigma_{68}(\nu)$, is the width of the
highest density interval for a $t$-distribution with scale parameter $\sigma =
1$ such that the probability contained in the interval is equal to that of a
1$\sigma$ interval for a normal distribution. This is given by
\begin{equation}
    \sigma_{68}(\nu) =
        \sqrt{
            \nu \left(
                \frac{1}{I^{-1}(\Phi_{\sigma=1};\frac\nu2, \frac12)} - 1
            \right)
        },
    \label{eqn:model.sigma_68}
\end{equation}
where $\Phi_{\sigma=1} = \text{erf}(2^{-1/2}) \approx 0.6827$ is the posterior
density contained within $\pm1\sigma$ of the mean for a normal distribution, and
$I^{-1}$ is the inverse of the regularized incomplete beta function, which we
define as
\begin{equation}
    I(x; a, b) = \frac{
        \int_0^x t^{a-1} (1 - t)^{b-1} {\rm d}t
    }{
        \int_0^1 t^{a-1} (1 - t)^{b-1} {\rm d}t
    }.
\end{equation}
This quantity is useful for defining a ``scale'' parameter that is less tightly
coupled to $\nu$ than the $\sigma$ parameter of the $t$-distribution.

\begin{figure}
	\includegraphics{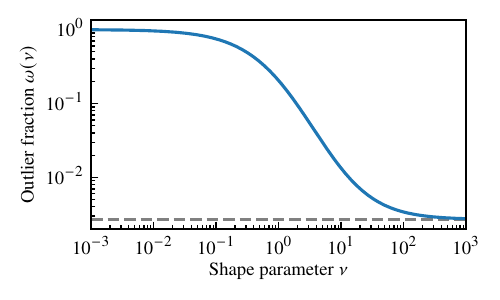}
    \caption{The outlier fraction, $\omega$, for different values of shape
    parameter, $\nu$. The outlier fraction of a normal distribution ($\approx
    2.70 \times 10^{-3}$) is indicated by the dashed grey line.}
    \label{fig:model.outlier_frac}
\end{figure}

The second quantity we use is an ``outlier fraction'' $\omega$, which is defined
as the fraction of points expected to lie more than $3\sigma$ from the
distribution mean $\mu$. For a $t$-distribution, this is given by
\begin{eqnarray}
    \omega(\nu)
    &=& P\left(
        \left|x - \mu \right| > 3 \sigma \;
        \middle| \;
        x \sim \studentt{\mu}{\sigma}
    \right) \label{eqn:model.outlier_frac}\\
    &=& 2 F_\nu \left(\mu - 3 \sigma \right),
\end{eqnarray}
where $F_\nu(x)$ is the cumulative distribution function for the Student's
$t$-distribution with shape parameter $\nu$. Figure~\ref{fig:model.outlier_frac}
shows the relationship between outlier fraction $\omega$ and shape parameter
$\nu$, with $\omega \rightarrow 2.70 \times 10^{-3} $ in the limit of a normal
distribution, i.e., as $\nu \rightarrow \infty$.  Under this definition
approximately 1 in 370 data-points would be outliers for data that follows a
normal distribution; for Cauchy-distributed data (i.e.\ $\nu = 1$), every fifth
data-point to be an outlier.

\subsection{Regression model}
\label{sec:formalism.model}

\begin{figure}
	\includegraphics[width=\columnwidth]{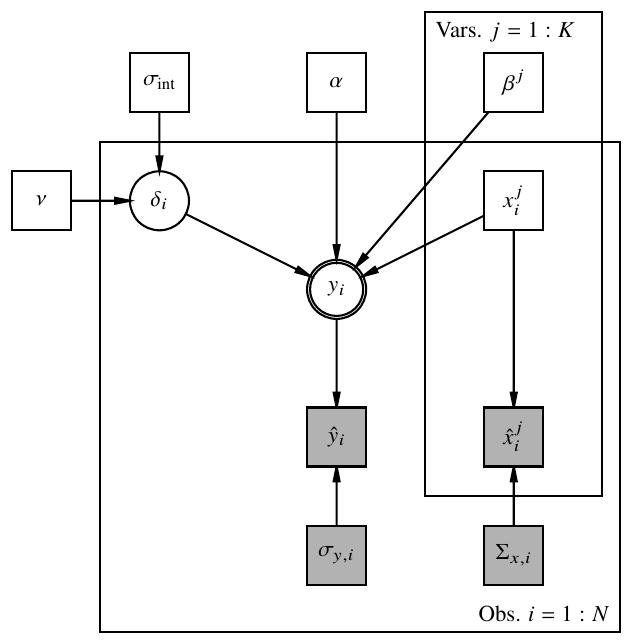}
    \caption{A directed acyclic graph representing the $t$-cup model for the
    function $f(\boldsymbol{x}; \boldsymbol{\theta}) = \alpha +
    \boldsymbol{\beta} \cdot \boldsymbol{x}_i$.}
    \label{fig:formalism.dag}
\end{figure}

Our regression model, represented as a directed acyclic graph in
Figure~\ref{fig:formalism.dag}, is specified here. We assume a linear
relationship between
$\{\boldsymbol{x}_i\}$ and $\{y_i\}$, i.e.\
\begin{equation}
    f(\boldsymbol{x}_i; \boldsymbol{\theta}_f) =
        \alpha + \boldsymbol{\beta} \cdot \boldsymbol{x}_i
\end{equation}
where $\{\alpha, \boldsymbol{\beta}\} \equiv \boldsymbol{\theta}_f$. In this
paper, we only consider this linear relationship, but this apparatus could be
used for other, non-linear relationships between variables with different
parameters.

Similarly to \citet{Kelly:2007}, we define a Bayesian hierarchical model to
reflect the nature of the regression problem. Firstly, we assume that we can
represent the probability distribution $\mathcal P_{\text{int}}$ with a
Student's $t$-distribution with shape parameter $\nu$ and scale parameter
$\sigma_{\text{int}}$ -- i.e.\
\begin{equation}
    \depvar \sim \studentt{\intercept + \covariate \indepvars}{\intscttr^2}.
\end{equation}
We use a Student's $t$-distribution here not because we believe that the data
intrinsically follows this distribution, but because the resulting model is
robust to model mis-specification -- such as if a galaxy were
to be an outlier from a particular relation as a result of a recent merger.

We assume that we can represent the probability distribution
$\mathcal P_{\text{obs}}$ as a normal distribution with scale parameter given by
the error associated with the measured quantity. These measurements are modelled
as
\begin{align}
    \obsindep \sim&\; \mathcal N\left({\indepvars}, \indepcov^2\right) \\
    \obsdep \sim&\; \mathcal N\left({\depvar}, \deperr^2\right).
\end{align}

We experimented with assuming $\mathcal P_{\text{obs}}$ to be $t$-distributed,
but found that the resultant model was difficult to sample as a result of its
geometry, and that posterior predictive checks often included large outliers and
did not resemble the datasets that gave rise to them.  Fortunately, for robust
inference the presence of a single heavy-tailed component is sufficient.

\subsection{Priors}
\label{sec:formalism.prior}

To ensure that our model is generically applicable to astronomical linear
regression, we ensure that both independent and dependent quantities are scaled
to have zero mean and unit variance. This allows us to set generic priors on the
scaled intercept, gradients and scatter $\{\tilde{\intercept},
\tilde{\covariate}, \tilde{\sigma}_{\text{int}}\}$ that do not require rescaling
for different units or datasets (though these priors can be revised to
incorporate prior information).

Our default priors on the regression parameters $\{\tilde{\intercept},
\tilde{\covariate}, \tilde{\sigma}_{\text{int}}\}$ are
\begin{align}
    \tilde{\intercept} \sim&\; \mathcal N(\mu = 0, \sigma^2 = 4), \\
    \tilde{\covariate} \sim&\; \mathcal N(\mu = 0, \sigma^2 = 4), \\
    \tilde{\sigma}_{68} \sim&\; \Gamma(\alpha = 1.1, \theta = 5).
\end{align}

The motivation behind the prior on $\tilde{\intercept}$ is that, as the data is
pre-scaled to have zero mean and the relationship between quantities is assumed
to be linear, we would expect the data to have zero intercept; we adopt a broad
prior by choosing a variance of 4. Similarly, the prior on $\tilde{\beta}$ has
been chosen because the pre-scaling of the data suggests a gradient between $-1$
and $1$, with variance chosen to give a weakly-informative prior. The prior on
$\tilde{\sigma}_{68}$ is informed by \citet{Chung:2013}, who demonstrate that a
prior with density at $\tilde{\sigma}_{\text{int}} = 0$ will lead to a
prediction of no intrinsic scatter in datasets where intrinsic scatter is
present; in addition, there is a reasonable physical argument that most
astrophysical processes will not produce objects with no population scatter. The
chosen prior balances this constraint with difficulties arising in testing as a
result of the pre-scaling, which led to insufficient prior density at small
levels of intrinsic scatter. We found that choosing a small value of the gamma
distribution power law slope placed more weight at smaller intrinsic scatter,
but required $\alpha > 1$ to ensure zero prior density at $\tilde{\sigma}_{68} =
0$; similarly, we found that a scale parameter of $\theta = 5$ ensured most of
the prior density fell in the range $0 < \tilde{\sigma}_{68} < 1$, which we
would expect because of the pre-scaling.

\begin{figure*}
	\includegraphics[width=\columnwidth]{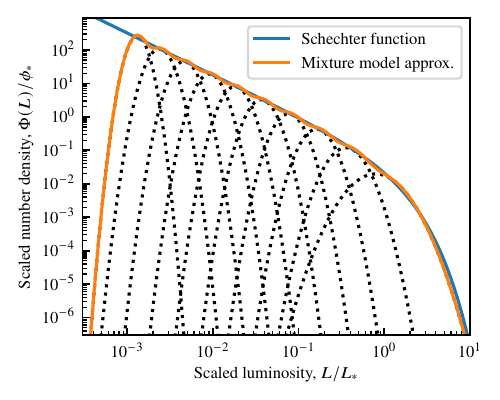}
	\includegraphics[width=\columnwidth]{
        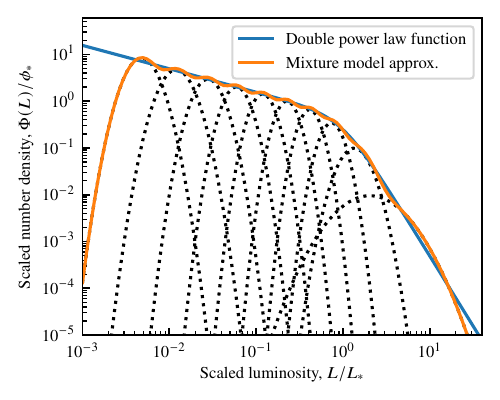
    }
    \caption{A mixture model approximation to a Schechter function (left) and a
    double power law function (right).}
    \label{fig:formalism.mixture-prior}
\end{figure*}

The prior on the latent values of the independent quantities $\{\indepvars\}$ is
a Gaussian mixture model prior, similar to that used in \citet{Kelly:2007}.
Although Gaussian mixture models may not seem suited to modelling distributions
that have strong power law gradients, previous work has found them to be more
than sufficient to reproduce standard astronomical distributions (e.g.\
\citealt{Blanton:2003, Kelly:2008} -- see also \citealt{Johnston:2011} for a
review of approaches). To illustrate the approximation, we fit a 10-component
Gaussian mixture model to two common population models in astronomy:
\begin{itemize}
    \item the \citet{Schechter:1976} function
    \begin{equation}
        \Phi(L; \alpha, L_*, \phi_*)
          = \phi_* \left(\frac{L}{L_*}\right)^\alpha
            \exp \left(-\frac{L}{L_*}\right)
    \end{equation}
    \item the double power law function (first used to model the quasar
    population in \citealt{Boyle:1987} and then presented explicitly in
    \citealt{Boyle:1988})
    \begin{equation}
        \Phi(L; \alpha, \beta, L_*, \phi_*)
          = \phi_* \left(\left(\frac{L}{L_*}\right)^\alpha
             + \left(\frac{L}{L_*}\right)^\beta\right)^{-1}.
    \end{equation}
\end{itemize}
Figure~\ref{fig:formalism.mixture-prior} shows that, while these functions are
well-approximated at the bright end of the luminosity function, if the sample is
flux-limited then the mixture model prior will fall off at the faint end when
the true population prior may increase. If sample selection effects are not
accounted for (as in this work), this mismatch could lead to Eddington bias
\citep{Eddington:1913} affecting the inferred parameters -- see
\citet{Andreon:2010} for an exploration of this effect. Similarly, using this
mixture prior for a flux-limited sample could lead to Malmquist bias
\citep{Malmquist:1922}, as brighter objects would be over-represented in the
sample, though this is unlikely to affect the inferred regression relation. In
such cases, another prior should be considered (and can be provided to the
$t$-cup package) -- for instance, if the luminosity function of the population
is well-characterized, this can be provided as an alternative prior and sidestep
these biases. An extension of the model that handles selection effects
\citep[e.g.][]{Kelly:2007} would eliminate these biases, but only in cases where
the selection function is well-defined; we defer a thorough treatment of a broad
range of selection functions to future work.

While \citeauthor{Kelly:2007} infers the complete prior as part
of the model, we found that such a prior led to sampling issues as a result of
the lack of identifiability inherent in a multicomponent Gaussian mixture model.
Therefore, we use extreme deconvolution \citep{Bovy:2011} to estimate the
parameters of a Gaussian mixture model approximating the latent distribution of
$\{\indepvars\}$, and use these parameters for the prior.

In principle, the exact choice of prior for $\nu$ is unimportant, as the model
is designed to be robust to model mis-specification; the only key element is
that $\nu$ can take on values which correspond to both normal and heavy-tailed
distributions. While difficulties in sampling $\nu$ according to a given prior
distribution would not necessarily invalidate the results of the model as long
as there is coverage at both small and large values, we use a prior that can be
efficiently sampled using HMC.  The prior we adopt on the shape parameter $\nu$
is
\begin{align}
    \nu \sim&\; \text{Inv-}\Gamma(4, 15),
\end{align}
which balances flexibility (such that the model could accommodate both heavily
leptokurtic distributions -- e.g.\ the Cauchy distribution -- and
normally-distributed data) and numerical considerations (as $\nu \rightarrow 0$,
the posterior geometry becomes highly curved, which leads to numerical issues;
as $\nu \rightarrow \infty$, different values of $\nu$ become rapidly
indistinguishable; both are down-weighted by this prior). The reasoning behind
this choice of prior is expanded upon in Appendix~\ref{sec:t-prior}.

\subsection{Asymptotic normality}
\label{sec:formalism.asymptotic}

\begin{figure}
    \includegraphics[
        width=\columnwidth,
        trim={0 0.3cm 0 0},
        clip
    ]{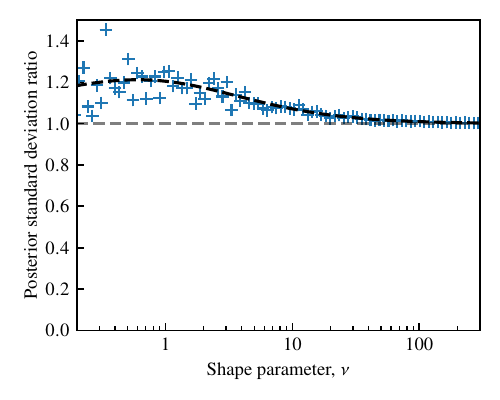}

    \caption{The ratio of the posterior standard deviation for a single location
    parameter when analysed with a $t$-distribution, compared to when analysed
    with a normal distribution. Each point represents a dataset with $N=100$
    datapoints. The black dashed line shows the expected ratio of posterior
    standard deviations as calculated using the Fisher information of the
    $t$-likelihood.}

    \label{fig:formalism.posterior-sd}
\end{figure}

One potential disadvantage of adopting a heavy-tailed likelihood in the
statistical model is that the resultant inferences would, in general, be less
constraining than under the assumption of a normal model (see, e.g., the
motivation behind the mixture model proposed in \citealt{Tak:2019}).  If there
are no outliers, the implication would be that the robust approach is inferior.
This effect is illustrated in Figure~\ref{fig:formalism.posterior-sd}, which
shows results for datasets of $N=100$ points drawn from a normal distribution
with zero mean and unit variance. The points show the ratio of the posterior
standard deviations for the mean when analyzed using a $t$-distribution to that
obtained using a normal distribution. These agree well with the approximate form
of this ratio obtained using the Fisher information of a $t$-distribution
likelihood (see Appendix~\ref{sec:fisher}). This ratio increases with decreasing
$\nu$ until $\nu \sim 0.6$, beyond which it falls again; this is a result of the
increasingly narrow peak of the $t$-distribution (see Figure~\ref{fig:model.t}).
While the uncertainty can be increased by up to $\sim20\%$ (in the Cauchy
regime, $\nu \approx 1$), we consider this an acceptable trade-off to reduce
bias in cases of model mis-specification.  This uncertainty trade-off can be
mitigated by a flexible approach in which $\nu$ is inferred (such as the one
presented here) as, if there are no outliers, higher values of $\nu$
corresponding to a normal distribution would be favoured.  The increased
numerical difficulty of treating $\nu$ as a parameter is justified by the
reduction in the uncertainty relative to fixed $\nu$ analyses.

This effect would be appreciable for the smallest datasets -- the posteriors
would have heavier tails than under a normal model -- but, in this case,
meaningful constraints are unlikely irrespective of the adopted model. For
larger datasets, of more than a few tens of points, these tails are effectively
multiplied out, typically leaving just a single region of high probability.
Moreover, for the regression problem considered here, the posterior in the
regression parameters satisfies the requirements for asymptotic normality
\citep[e.g.][]{Ghosh_etal:2006}, so the core of the posterior is Gaussian in
form, just as would be the case under a normal model.  (There is not even a
significant numerical cost as the posterior has to be evaluated by sampling
anyway.)

\subsection{Implementation}
\label{sec:formalism.implementation}

The model described in Section~\ref{sec:formalism.model} is implemented in
NumPyro \citep{NumPyro, Pyro}, which is used to draw samples via a HMC No U-Turn
Sampler. For the most part, we implement the model as presented previously;
however, we reparametrize the Student's $t$-distribution using a mixture model
of normal distributions \citep[e.g.][]{Stan}. We found this to be necessary for
any model with non-negligible prior density below $\nu \lesssim 1$, as HMC
struggles with the extreme gradients that would be present if the model were
implemented using Student $t$-distributions directly \citep{Neal:2003,
Betancourt:2013}.  The implementation is packaged as $t$-cup, available as a
Python package\footnotemark.

For the purposes of comparison, we have also implemented a model that mirrors
the above structure, but uses normal distributions at each stage -- we refer to
this as $n$-cup.

\footnotetext{\url{https://github.com/wm1995/tcup}}

\section{Validation}
\label{sec:methods}

In this section, we outline the methods used to validate the model set out in
Section~\ref{sec:formalism}. We run two types of tests to validate the
performance of $t$-cup under different types of model mis-specification:
simulation-based calibration tests \citep{Cook:2006, Talts:2018}; and fixed
value calibration tests.

\subsection{Simulation-based calibration}
\label{sec:methods.sbc}

Simulation-based calibration tests \citep{Cook:2006, Talts:2018} are used to
diagnose sampling issues. Parameters of interest $\theta_0$ are drawn from the
model prior $\pi(\theta)$, and a dataset is drawn following the prescription of
the model. We can then run a single MCMC chain on this dataset until we have
drawn $L$ independent samples $\{\tilde{\theta}_{i}\}$ (see \citet{Talts:2018}
for a discussion of independence criteria). We then compute the rank statistic
\begin{equation}
    r(\theta_0, \tilde{\boldsymbol{\theta}}_{\text{MCMC}})
        = \sum_{i = 1}^{L} \mathbb I (\tilde{\theta}_i < \theta_0),
\end{equation}
where $\mathbb{I}(\cdot)$ is the indicator function which evaluates to 1 if its
(logical) argument is true and 0 if it is false.  If the generative model
matches the sampling model and the sampling algorithm correctly draws from the
posterior, the rank statistic will be uniformly distributed across the integers
$[0, L]$. We can, therefore, repeat this process many times to build a histogram
of rank statistics; any shortfalls in the sampling algorithm will manifest as
deviations from uniformity.

While this procedure has been developed to diagnose sampling issues, we can also
use the method to build a heuristic measure of how robust a model is to
mis-specification. If the generative model and the sampling model no longer
match, we would expect deviations from uniformity in the rank statistic
histogram. We can compare how well different sampling models deal with model
mis-specifcation by building the rank statistic histogram and assessing the
deviation from uniformity.

As our priors are defined in a scaled space, our simulation-based calibration
tests are also conducted in this scaled space; in this way, we are solely
testing the performance of the MCMC models, and not of the scaling before
fitting the models.

\subsection{Fixed-value calibration}
\label{sec:methods.fixed}

For our fixed-value calibration tests, we simulate datasets from known models
with true fixed values for our regression parameters. The purpose of these tests
is to demonstrate accurate recovery of these values, and to compare with the
results given when using normal distributions.

We generate multiple datasets with the same ground-truth parameters to verify
the results across multiple runs, producing plots of the composite cumulative
distribution function across each dataset. In each instance, full specifications
of the data models are given in Appendix~\ref{sec:data-models}.

We aim to do a full end-to-end test of our inference pipeline in the fixed-value
calibration tests; therefore, these datasets are generated in the unscaled space
and are a test not only of the sampler but also of the scaling.

\section{Results on simulated datasets}
\label{sec:results}

In the previous section, we proposed a general-purpose, robust statistical model
for linear regression; in this section, we investigate the performance of the
model on a series of simulated datasets with known parameters. Code that
reproduces the datasets in this section is available online\footnotemark.

\footnotetext{\url{https://github.com/wm1995/tcup-paper}}

\subsection{\textit{t}-distributed data}
\label{sec:results.t}

\begin{figure*}
    \includegraphics[width=\textwidth]{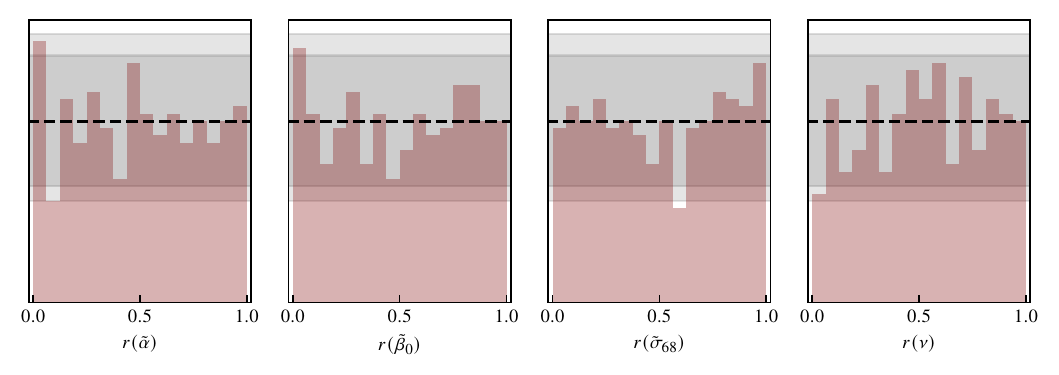}
    \caption{SBC runs for the t distribution under $t$-cup. If the inference
    procedure is working as expected, the histograms for each parameter should
    be distributed uniformly (as indicated by the black dashed line). The dark
    (light) grey regions correspond to the 94\% (98\%) confidence interval of
    uniformity (i.e.\ we expect one histogram bin per panel (figure) to lie
    outside of this range).}
    \label{fig:results.t.sbc}
\end{figure*}

For our first test, we start with a dataset that matches our model perfectly
(i.e.\ $\mathcal P_{\text{int}} = t_\nu, \mathcal P_{\text{obs}} =
\mathcal{N}$), and choose $K=1$ independent variables. The simulation-based
calibration tests indicate that the rank statistic is consistent with being
uniformly distributed (see Figure~\ref{fig:results.t.sbc}), and, therefore, that
our inference procedure is working as expected.

\begin{figure}
    \includegraphics[width=\columnwidth]{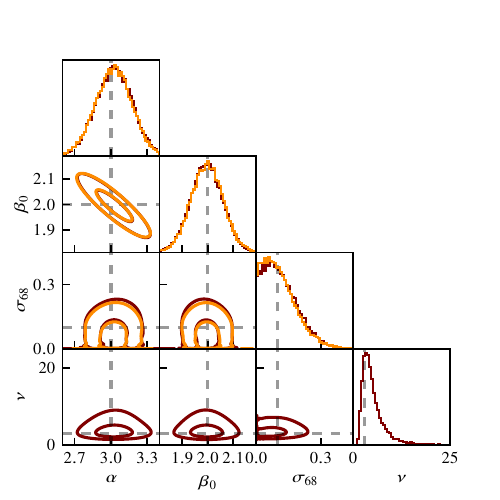}
    \caption{The posterior for each of the regression coefficients and the shape
    parameter $\nu$ for the \textit{t}-distributed dataset, with ground-truth
    values indicated by the black dashed lines. Constraints from $t$-cup (red)
    are compared with those derived with $\nu$ fixed to the true value (orange).
    Contours indicate 39.3\% and 86.5\% highest posterior density regions,
    corresponding to 1$\sigma$ and 2$\sigma$ contours for a bivariate normal
    distribution.}
    \label{fig:results.t.corner}
\end{figure}

For our fixed-value tests, we drew $N = 20$ datapoints with $(\alpha, \beta,
\sigma_{68}, \nu) = (3, 2, 0.1, 3)$; the full model used to generate the data is
specified in Appendix~\ref{sec:data-models.t}. The chosen value of $\nu = 3$
corresponds to an outlier fraction of $\omega(\nu = 3) \approx 5.8 \%$.  As
shown in Figure~\ref{fig:results.t.corner}, we recover the values of the
parameters that were used to generate the dataset; these values are consistent
with a model where the shape parameter is fixed to $\nu = 3$.

We then generated 400 datasets from the same ground-truth parameters, and ran
MCMC against each dataset.
95\% highest posterior density credible intervals were constructed for each run;
the true parameter values were contained in these credible intervals $97\%$,
$96\%$ and $98\%$ of the time for the intercept $\alpha$, the slope $\beta$ and
the intrinsic scatter $\sigma_{68}$. For the nuisance parameter $\nu$, the
credible intervals contained the true parameter value across all runs; this may
be caused by this parameter being only weakly constrained for these datasets.

\subsection{Normally-distributed data}
\label{sec:results.outlier}

In this test, we compare $t$-cup with an equivalent model that employs normal
distributions to check that:
\begin{enumerate}
    \item $t$-cup reduces to a normal model in the absence of outliers
    \item $t$-cup gives less biased results when an extreme outlier is present.
\end{enumerate}

\begin{figure*}
    \includegraphics[width=\textwidth]{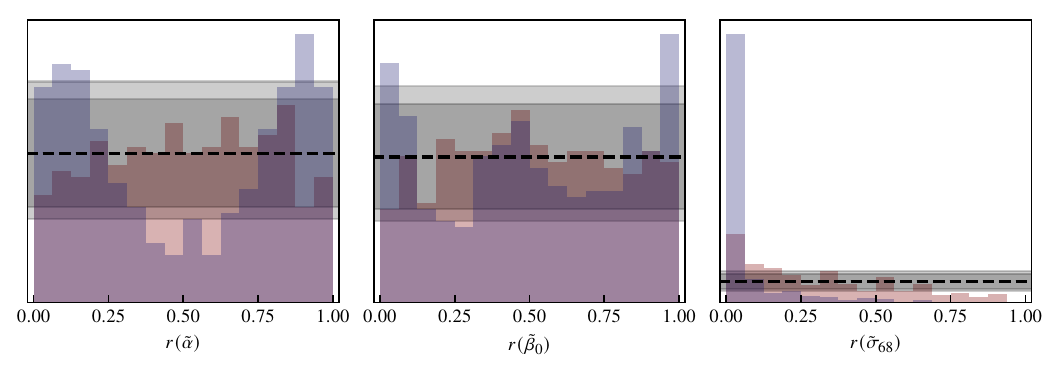}
    \caption{A rank statistic histogram for 400 simulation-based calibration
    runs for the outlier dataset under $n$-cup (blue) and $t$-cup (red). If
    there were a perfect match between the generative model and the inference
    model, the histograms for each parameter will be distributed uniformly (as
    indicated by the black dashed line). The dark (light) grey regions
    correspond to the 94\% (98\%) confidence interval of uniformity (i.e.\ we
    expect one histogram bin per panel (figure) to lie outside of this range).}
    \label{fig:results.outlier.sbc}
\end{figure*}

We conducted a simulation-based calibration test, where datasets were generated
using normal distributions throughout (i.e.\ $\mathcal P_{\text{int}} = \mathcal
P_{\text{obs}} = \mathcal{N}$) and with a single independent variable (i.e.\ $K
= 1$). The results (shown in Figure~\ref{fig:results.outlier.sbc}) indicate
that, while the estimates of the intercept and slope are significantly biased
when analysed by the normal model (as indicated by the rank statistic's
distribution deviating from normal), the estimates derived by $t$-cup are
unbiased. Additionally, while both $n$-cup and $t$-cup systematically
overestimate the intrinsic scatter $\sigma_{68}$ (as defined in
Equation~\ref{eqn:model.sigma_68}), the bias is much smaller for $t$-cup than
for $n$-cup.

A dataset of $N = 12$ points was generated with $(\alpha, \beta, \sigma_{\rm
int}) = (3, 2, 0.2)$, and one of the points was modified to be a
$\sim20\,\sigma$ outlier (the full generative model is given in
Appendix~\ref{sec:data-models.outlier}). While such an extreme outlier could
easily be identified and removed, the purpose here is to demonstrate that
$t$-cup does not require this to obtain sensible inferences.

\begin{figure*}
    \includegraphics[width=\linewidth]{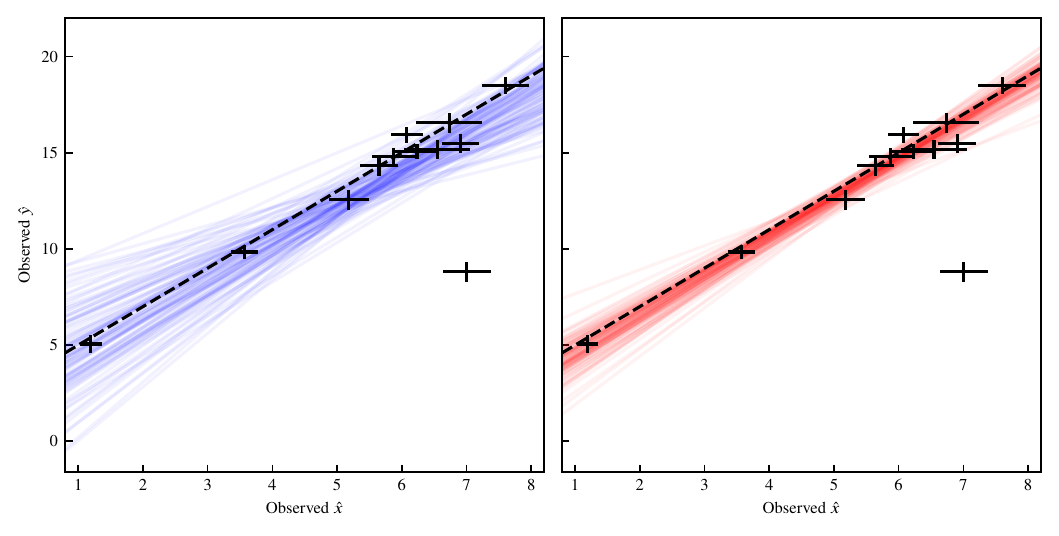}
    \caption{100 draws from the posterior of regression lines from
    the normal model (left panel, blue) and $t$-cup (right panel, red). The
    dataset is illustrated by the black points, with the ground-truth regression
    line illustrated by the black dashed line.}
    \label{fig:results.outlier.regression}
\end{figure*}

\begin{figure}
    \includegraphics[width=\columnwidth]{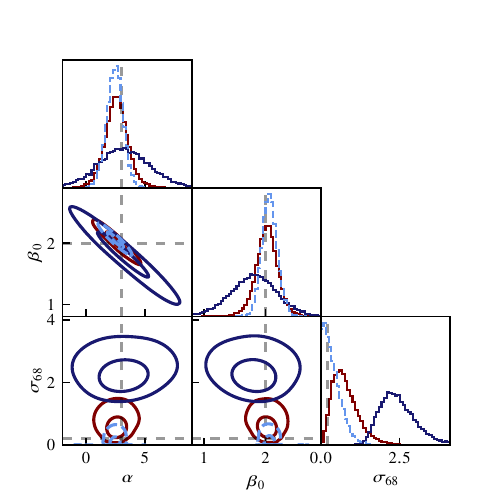}
    \caption{The posterior for each of the regression coefficients under the
    normal model with the outlier included (dark blue, solid) and excluded
    (light blue, dashed), and for $t$-cup with the outlier included (dark red,
    solid). Ground-truth values are indicated by the black dashed lines.}
    \label{fig:results.outlier.corner}
\end{figure}

\begin{figure}
    \includegraphics[width=\columnwidth]{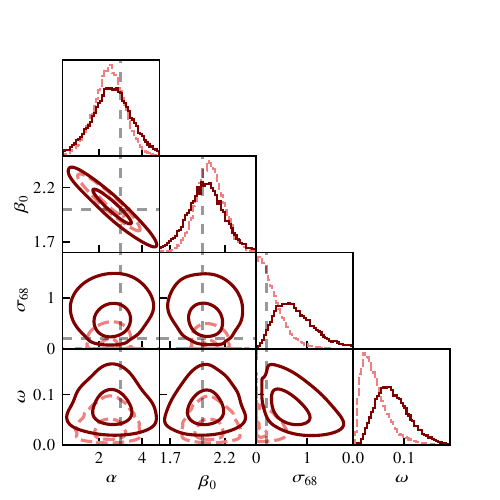}
    \caption{The posterior for each of the regression coefficients and the shape
    parameter $\nu$ under the $t$-cup model with the outlier included (dark red,
    solid) and excluded (light red, dashed). Ground-truth values are indicated
    by the black dashed lines.}
    \label{fig:results.outlier.tcup}
\end{figure}

Figure~\ref{fig:results.outlier.regression} illustrates how the estimates for
the true parameter models are biased in the normal model, but less affected in
the $t$-cup model.

In Figure~\ref{fig:results.outlier.corner}, we compare the constraints on
parameters derived under the normal model (including and excluding the outlier
from the dataset), and the $t$-cup model (including the outlier only). The
constraints from the $t$-cup model including the outlier are consistent with
those derived under the normal model when the outlier is excluded, obviating the
need to remove the outlier manually. While this outlier is particularly extreme,
this example illustrates the utility of the $t$-cup model in datasets with
outliers.

For completeness, in Figure~\ref{fig:results.outlier.tcup} we illustrate that
the $t$-cup model recovers consistent constraints regardless of whether the
outlier is included or excluded.

\begin{figure*}
    \includegraphics[
        width=\textwidth,
        trim={0 0.3cm 0 0},
        clip
    ]{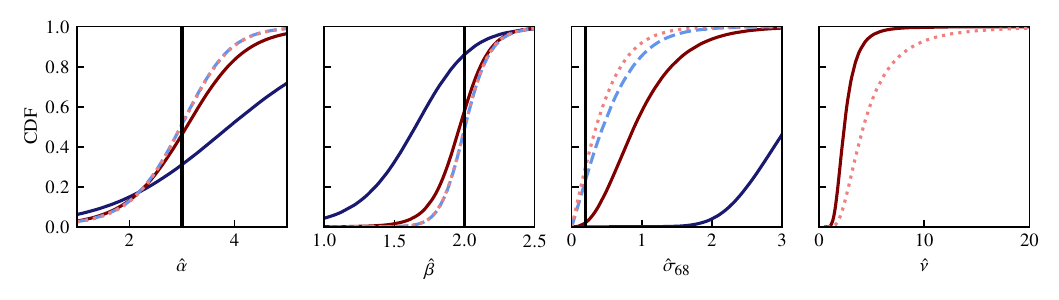}
    \caption{The combined cumulative distribution function of the regression
    parameters for 400 normal datasets with an outlier under the normal model
    (dark blue, solid) and $t$-cup model (dark red, solid), and with the outlier
    removed for the normal model (light blue, dashed) and $t$-cup model (light
    red, dashed).}
    \label{fig:results.outlier.map}
\end{figure*}

We then generated 400 datasets using the same procedure, and combined posterior
samples from each run to build an effective cumulative distribution function
across all runs for both the normal model and for $t$-cup. The results
(illustrated in Figure~\ref{fig:results.outlier.map}) indicate that constraints
under $t$-cup are significantly less biased than those calculated under the
normal model.

%

\subsection{Mis-specified observational error}

\begin{figure*}
    \includegraphics[width=\textwidth]{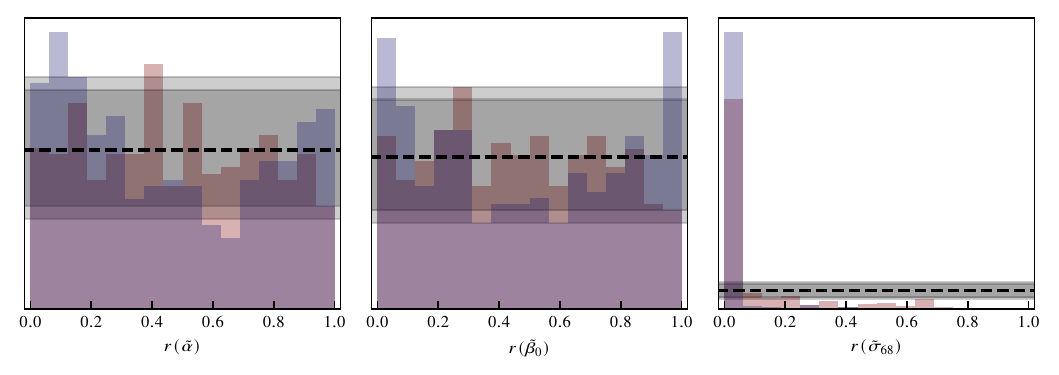}
    \caption{A rank statistic histogram for 400 simulation-based calibration
    runs for the heavy-tailed observation dataset under $n$-cup (blue) and
    $t$-cup (red). If there were a perfect match between the generative model
    and the inference model, the histograms for each parameter would be
    distributed uniformly (as indicated by the black dashed line). The dark
    (light) grey regions correspond to the 94\% (98\%) confidence interval of
    uniformity (i.e.\ we expect one histogram bin per panel (figure) to lie
    outside of this range).}
    \label{fig:results.obs.sbc}
\end{figure*}

This test explores how $t$-cup performs when the distribution $\mathcal P_{\rm
obs}$ is mis-specified. We conducted a simulation-based calibration test where
$\mathcal P_{\rm int}$ is a normal distribution, and used $\mathcal P_{{\rm
obs}, x} = {\rm Laplace}$, $\mathcal P_{{\rm obs}, y} = {\rm Cauchy}$ for
observational error. The results (see Figure~\ref{fig:results.obs.sbc}) show
that $t$-cup is able to recover both the intercept and slope without bias, while
$n$-cup produces biased estimates of both. As with the outlier case above, both
$t$-cup and $n$-cup show significant bias in recovering the intrinsic scatter,
but the bias is smaller for $t$-cup.

\subsection{Two-dimensional normal mixture model with outliers}
\label{sec:results.gmm}

This test is designed to further explore how $t$-cup performs when the model is
misspecified. In this case, we are looking at a normally-distributed population
which has a 10\% contamination rate with another normally-distributed population
of the same mean but 10 times the standard deviation. Equivalently, this test
can be thought of as investigating how the model performs when there is a
significant fraction of outliers.

The intrinsic scatter distribution $\mathcal P_{\rm int}$ is a mixture of two
normal distributions with zero mean; 90\% of points are drawn from a core
distribution with standard deviation $\sigma_\textrm{int}$, and 10\% of points
are drawn from an outlier distribution with standard deviation $10
\sigma_\textrm{int}$. The observation distribution $\mathcal P_{\rm obs}$ is a
normal distribution. For fixed-value tests, the true values of the regression
parameters were fixed to $(\alpha, \beta_0, \beta_1, \sigma_{\rm int}) = (2,
3, 1, 0.4)$. The full generative model for the fixed-value tests is given in
Appendix~\ref{sec:data-models.gmm}.

\begin{figure}
    \includegraphics[width=\columnwidth]{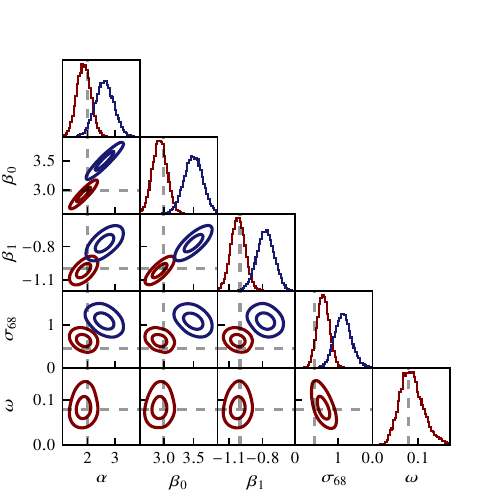}
    \caption{The posterior for each of the regression coefficients and the
    outlier fraction $\omega$ for the normal mixture model dataset for $n$-cup
    (blue) and $t$-cup (red), with ground-truth values indicated by the grey
    dashed lines.}
    \label{fig:results.gmm.corner}
\end{figure}

The results (see Figure~\ref{fig:results.gmm.corner}) show that, while the
constraints from the normal model are significantly biased for this generative
distribution, $t$-cup is able to recover the correct parameter values.

\subsection{Laplace-distributed data}
\label{sec:results.laplace}

As the underlying sampling distributions can never be known with certainty it is
important to test the method on data generated on multiple distributions that do
not fall into the family of $t$-distributions. Here we use a Laplace
distribution as $\mathcal P_{\rm int}$, giving another example of performance
under explicit model mis-specification. The Laplace distribution has probability
density
\begin{equation}
    {\rm Laplace}\left(x; \mu, b\right) =
        \frac1{2b} \exp\left(-\frac{\left|x - \mu\right|}{b}\right),
\end{equation}
with scale parameter $b$ related to $\sigma_{68}$ as $b(\sigma_{68}) =
- \sigma_{68} / \ln \left( 1 - {\rm erf} (2^{-1/2})\right) \approx 0.87\,
\sigma_{68}$, where $\sigma_{68}$ is defined in
Equation~\ref{eqn:model.sigma_68}.

We generate $N = 25$ datapoints from a model with Laplacian intrinsic scatter
(i.e.\ $\mathcal P_{\rm int} = \mathrm{Laplace}$), normally distributed
observation noise (i.e.\ $\mathcal P_{\rm obs} = \mathcal N$), and $K = 1$
independent variables. For the fixed-value tests, we set $(\alpha, \beta, b) =
(-1, 0.8, 0.2)$. The full generative model can be found in
Appendix~\ref{sec:data-models.laplace}.

\begin{figure}
    \includegraphics[width=\columnwidth]{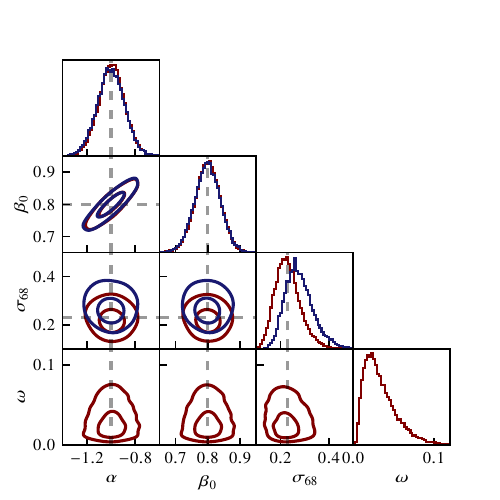}
    \caption{The posterior for each of the regression coefficients and the shape
    parameter $\nu$ for the Laplace-distributed dataset under the normal model
    (blue) and $t$-cup (red), with ground-truth values indicated by the black
    dashed lines.}
    \label{fig:results.laplace.corner}
\end{figure}

\begin{figure*}
    \includegraphics[width=\textwidth, trim={0 0.3cm 0 0}, clip]{
        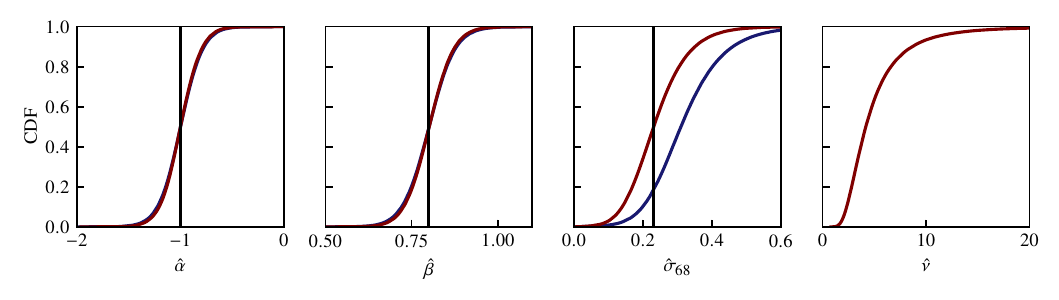
    }
    \caption{The combined cumulative distribution function of the regression
    parameters for 400 Laplace datasets under the normal model
    (blue) and $t$-cup model (red).}
    \label{fig:results.laplace.map}
\end{figure*}

As we can see in Figure~\ref{fig:results.laplace.corner}, while both $n$-cup and
$t$-cup successfully recover the gradient and intercept that were used to
generate the dataset, $n$-cup overpredicts the true intrinsic scatter when
$t$-cup constrains it accurately. To confirm this, we generated 400 datasets
using the same procedure, and analysed each with both $n$-cup and $t$-cup. The
results (illustrated in Figure~\ref{fig:results.laplace.map}) show the same
pattern -- that $t$-cup is able to accurately constrain the intrinsic scatter,
while the estimate from the normal model is biased high.

\section{Demonstration on real data}
\label{sec:real-world}

Having seen $t$-cup's performance on simulated data in the previous section, we
now compare the performance of the $t$-cup model with a generic astronomical
Bayesian linear regression model \citep[\textsc{linmix\_err};][]{Kelly:2007} and
a tailored approach using $t$-distributions \citep{Park:2017}.

\subsection{\textsc{linmix\_err}}

\begin{figure*}
    \includegraphics[width=\linewidth]{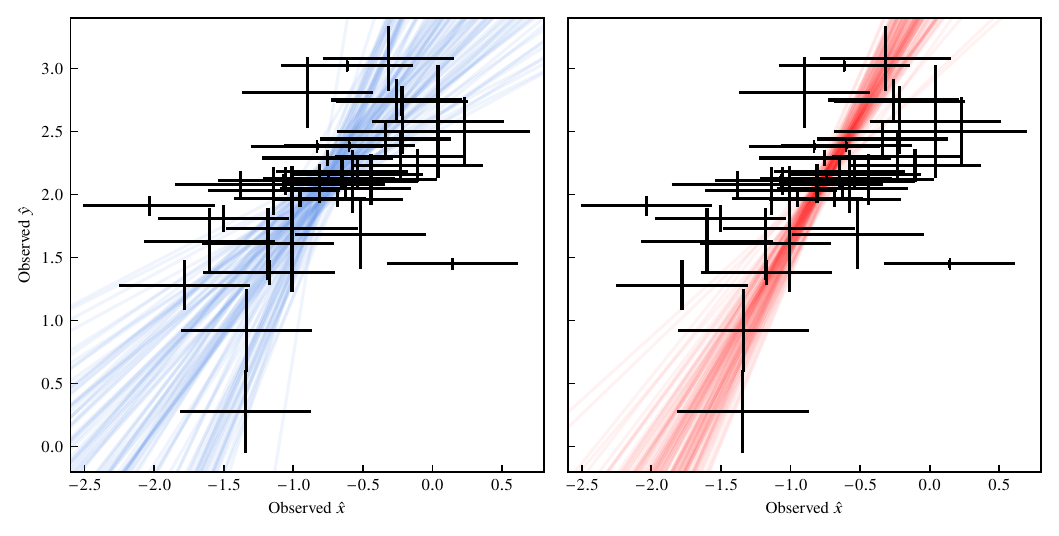}
    \caption{100 draws from the posterior of regression lines from
    \textsc{linmix\_err} (left panel, blue) and $t$-cup (right panel, red) for
    the \citet{Kelly:2007} dataset.}
    \label{fig:real-world.kelly.regression}
\end{figure*}

We use the same dataset that is used in Section~8 of \citet{Kelly:2007} -- a
dataset of $N = 39$ quasars with measured Eddington ratio $L / L_{\text{Edd}}$
and X-ray spectral index $\Gamma$. Performance is compared with a Python
implementation\footnotemark of the original \textsc{linmix\_err} paper proposed
by \citeauthor{Kelly:2007} -- see Figure~\ref{fig:real-world.kelly.regression}.

\footnotetext{\url{https://github.com/jmeyers314/linmix}}

\begin{table}
	\centering
	\caption{Estimates for the intercept, slope, and intrinsic scatter inferred
	for the relationship between Eddington ratio $L / L_{\text{Edd}}$ and X-ray
	spectral index $\Gamma$ for the sample of quasars analysed by
	\citet{Kelly:2007}. The parameter estimates reported in \citet{Kelly:2007}
	are the posterior median and ``a robust estimate of the standard
	deviation''; for linmix and $t$-cup, we report the posterior median and an
	estimate of the standard deviation as $\sigma = 1.4826 \, \text{MAD}$, where
	MAD is the median absolute deviation.}
	\label{tab:real-world.kelly.params}
	\begin{tabular}{lccc} 
                               & \citet{Kelly:2007} & linmix   & $t$-cup       \\
    Intercept $\alpha$         & 3.12 $\pm$ 0.41 & 3.18 $\pm$ 0.48 & 3.62 $\pm$ 0.26 \\
    Slope $\beta$              & 1.35 $\pm$ 0.54 & 1.40 $\pm$ 0.60 & 2.00 $\pm$ 0.33 \\
    Int. scatter $\sigma_{68}$ & 0.26 $\pm$ 0.11 & 0.25 $\pm$ 0.12 & 0.08 $\pm$ 0.07 \\
    Outlier fraction $\omega$  &      ---      &      ---      & 0.04 $\pm$ 0.03 \\
\end{tabular}
\end{table}

\begin{figure}
    \includegraphics[width=\columnwidth]{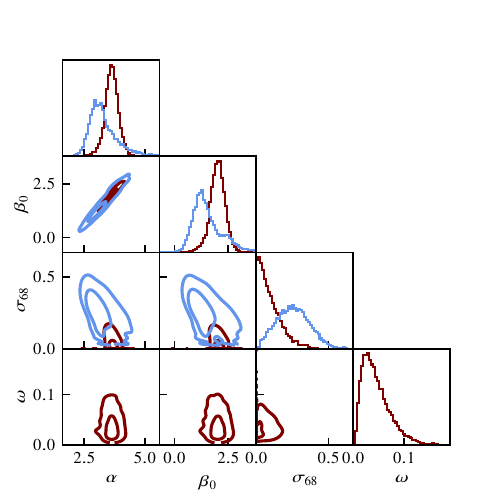}
    \caption{The posterior distributions for the data from \citet{Kelly:2007}
    under the linmix (blue) and $t$-cup (red) models.}
    \label{fig:real-world.kelly.corner}
\end{figure}

While the parameter estimates are broadly consistent (see
Table~\ref{tab:real-world.kelly.params}), the posterior distributions in
Figure~\ref{fig:real-world.kelly.corner} can be seen to differ, with the
posterior inferred by \textsc{linmix} being more diffuse than that inferred by
$t$-cup.  The tighter constraints of $t$-cup suggest that the inference
presented by \citet{Kelly:2007} may be biased by the enforced assumption of
normally-distributed data, which may not be accurate. The explicit assumption of
normality in \textsc{linmix} is in tension with the outlier fraction estimated
by $t$-cup (68\% CI $\omega = 0.05$\raisebox{0.5ex}{\tiny$^{+0.04}_{-0.01}$}).
This showcases that analysing data with robust procedures gives materially
different answers on real-world data, and highlights the importance of carefully
considering the implications when assuming normality.

\subsection[Park et al. (2017)]{\citet{Park:2017}}

\begin{figure}
    \includegraphics[width=\columnwidth]{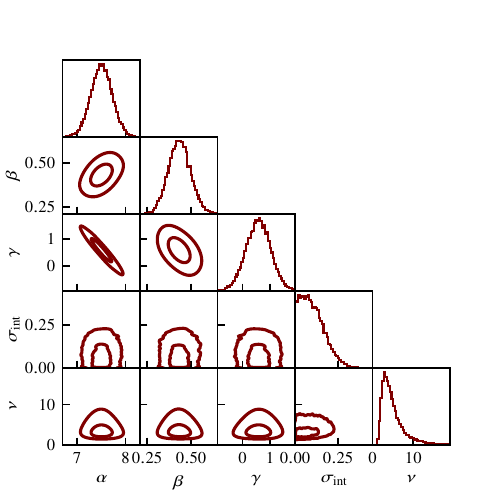}
    \caption{The posterior distributions for the data from \citet{Park:2017}
    under the $t$-cup model. This figure is directly comparable with Figure~9
    from \citet{Park:2017}.}
    \label{fig:real-world.park.corner}
\end{figure}

\citet{Park:2017} presents a bespoke $t$-distribution-based linear regression
model for estimating SMBH mass; this allows us to compare the results of their
bespoke model with our generic one. Their dataset consists of $N = 31$ AGN with
reverberation-mapped mass estimates, $M_{\text{BH}}$, \ion{C}{iv} emission line
width, $\Delta V$, and continuum luminosities, $\lambda L_{\lambda}$. Three
different \ion{C}{iv} line width measurements are compared in \citet{Park:2017}:
the full-width at half maximum (FWHM), the line dispersion, $\sigma_l$, and the
median absolute deviation. This dataset is used to fit the regression relation
\begin{equation}
    \log_{10}\!\left( \frac{M_{\text{BH}}}{10^8 M_\odot} \right)\! =
        \alpha +
        \beta \log_{10}\! \left( \frac{\lambda L_{\lambda}}{10^{44} \text{erg s}^{-1}} \right) +
        \gamma \log_{10}\!\left( \frac{\Delta V}{10^3 \text{km s}^{-1}} \right).
        \label{eqn:real-world.park}
\end{equation}
The $t$-cup posterior for parameters $\alpha$, $\beta$ and $\gamma$, as well as
intrinsic scatter $\sigma_{68}$ and shape parameter $\nu$, is shown in
Figure~\ref{fig:real-world.park.corner}.

The constraints on the intercept, $\alpha$, and slopes, $\beta$ and $\gamma$,
are consistent with those derived by \citet{Park:2017} for all three measures of
emission line width -- see Table~\ref{tab:real-world.park.params}.
\citet{Park:2017} find a systematically higher intrinsic scatter than that
obtained using $t$-cup; however, these figures cannot be compared directly, as
we do not know the exact statistical model used by \citet{Park:2017}.

\begin{table*}
	\centering
	\caption{A comparison between the constraints on the parameters in
	Equation~\ref{eqn:real-world.park} derived by \citet{Park:2017} and using
	$t$-cup.}
	\label{tab:real-world.park.params}
	\begin{tabular}{lcc|cc|cc} 
                          & \multicolumn{2}{c}{FWHM} & \multicolumn{2}{c}{$\sigma_l$} & \multicolumn{2}{c}{MAD} \\
                          & \citet{Park:2017} & $t$-cup & \citet{Park:2017} & $t$-cup & \citet{Park:2017} & $t$-cup \\
    Intercept $\alpha$    & $7.54$\raisebox{0.5ex}{\tiny$^{+0.26}_{-0.27}$} & $7.51$\raisebox{0.5ex}{\tiny$^{+0.22}_{-0.22}$} & $6.90$\raisebox{0.5ex}{\tiny$^{+0.35}_{-0.34}$} & $6.91$\raisebox{0.5ex}{\tiny$^{+0.30}_{-0.31}$} & $7.15$\raisebox{0.5ex}{\tiny$^{+0.24}_{-0.25}$} & $7.15$\raisebox{0.5ex}{\tiny$^{+0.22}_{-0.22}$} \\
    Slope $\beta$         & $0.45$\raisebox{0.5ex}{\tiny$^{+0.08}_{-0.08}$} & $0.43$\raisebox{0.5ex}{\tiny$^{+0.06}_{-0.06}$} & $0.44$\raisebox{0.5ex}{\tiny$^{+0.07}_{-0.07}$} & $0.43$\raisebox{0.5ex}{\tiny$^{+0.05}_{-0.06}$} & $0.42$\raisebox{0.5ex}{\tiny$^{+0.07}_{-0.07}$} & $0.42$\raisebox{0.5ex}{\tiny$^{+0.05}_{-0.06}$} \\
    Slope $\gamma$        & $0.50$\raisebox{0.5ex}{\tiny$^{+0.55}_{-0.53}$} & $0.58$\raisebox{0.5ex}{\tiny$^{+0.44}_{-0.45}$} & $1.66$\raisebox{0.5ex}{\tiny$^{+0.65}_{-0.66}$} & $1.66$\raisebox{0.5ex}{\tiny$^{+0.57}_{-0.58}$} & $1.65$\raisebox{0.5ex}{\tiny$^{+0.61}_{-0.62}$} & $1.65$\raisebox{0.5ex}{\tiny$^{+0.55}_{-0.55}$} \\
    Int. scatter $\sigma$ & $0.16$\raisebox{0.5ex}{\tiny$^{+0.10}_{-0.08}$} & $0.10$\raisebox{0.5ex}{\tiny$^{+0.03}_{-0.10}$} & $0.12$\raisebox{0.5ex}{\tiny$^{+0.09}_{-0.06}$} & $0.08$\raisebox{0.5ex}{\tiny$^{+0.02}_{-0.08}$} & $0.12$\raisebox{0.5ex}{\tiny$^{+0.09}_{-0.06}$} & $0.07$\raisebox{0.5ex}{\tiny$^{+0.02}_{-0.07}$} \\
    \end{tabular}
\end{table*}

\section{Conclusions}
\label{sec:conclusion}

We have presented a general-purpose approach to linear regression, implemented
as $t$-cup, that is robust to model mis-specification, with a model laid out in
Section~\ref{sec:formalism}. In Section~\ref{sec:results}, we demonstrated that
the model recovers constraints consistent with those used to generate the
datasets, including in cases where there is a mismatch between the generative
model used to create the dataset and our regression model. In
Section~\ref{sec:real-world}, we compared the method to \citet{Kelly:2007} and
\citet{Park:2017} on real-world data, illustrating that the models derive
consistent constraints.

It may be fruitful to re-examine some of the priors assumed in the $t$-cup
model; while we focused on producing a model that was applicable to a large
range of datasets, $t$-cup predicted lower intrinsic scatter than both
\citet{Kelly:2007} and \citet{Park:2017}. While results between the three are
not directly comparable as a result of the different assumptions in each model,
this difference could suggest that the prior on $\sigma_{68}$ has too much
density near $\sigma_{68} = 0$. In addition, setting the prior on
$\{\mathbf{x}_i\}$ using extreme deconvolution, while effective, is not
theoretically motivated; a prior similar to that presented in
\citet{Bartlett:2023} may be more appropriate in this case.

In our next paper, we will apply the robust inference techniques presented here
to exploring the single-epoch mass estimators used in estimating the masses of
high-redshift quasars.

\section*{Acknowledgements}

The authors would like to thank Alan Heavens, Stephen Feeney, Boris Leistedt,
Jonathan Pritchard, Justin Alsing, Tamas Budavari and Chad Schafer for useful
conversations, and Tianning Lyu for a typographical correction. We also thank
the anonymous referees for their comments and suggestions. WM is supported by
STFC grant ST/T506151/1.

\section*{Data Availability}

The data in this paper is sourced from \citet{Kelly:2007} and \citet{Park:2017}.
Scripts to generate the simulated datasets used to validate the model in
Section~\ref{sec:results} are available from the GitHub repository for this
paper, at \url{https://github.com/wm1995/tcup-paper}.


\bibliographystyle{rasti}
\bibliography{references}



\appendix

\section{Prior choice for shape parameter}
\label{sec:t-prior}

As our model relies on Student's $t$-distributions, we review notation and
priors used by previous works and justify our reasoning for our prior choice.
One approach is to adopt a fixed value of $\nu$, which is equivalent to setting
a Dirac delta function prior: a common choice is $\nu = 4$
\citep[e.g.][]{Berger:1994, Gelman:2013}.  Another approach is to adopt a more
flexible approach by allowing $\nu$ to vary \citep[e.g.][]{Juarez:2010,
Gelman:2013, Ding:2014, Park:2017, Feeney:2018}.

\begin{figure}
	\includegraphics[width=\columnwidth]{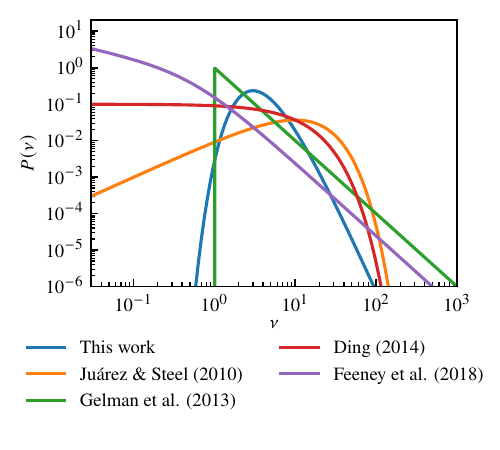}
    \caption{A selection of different priors on $\nu$ in works that have used
    $t$-distributions.}
    \label{fig:priors.pdf}
\end{figure}

\begin{figure}
	\includegraphics[width=\columnwidth]{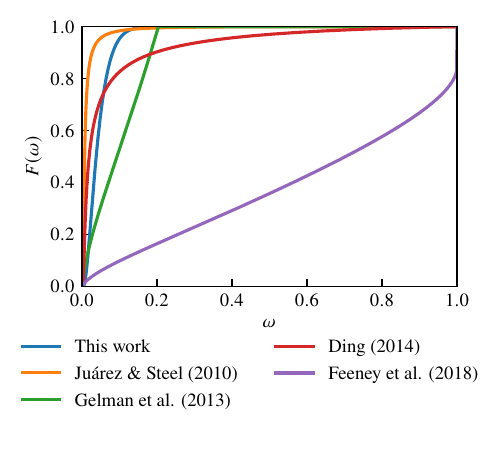}
    \caption{The CDF of priors on $\nu$ used in works that have used
    $t$-distributions. The priors are expressed in terms of outlier fraction,
    $\omega$, as defined in Equation~\ref{eqn:model.outlier_frac}.}
    \label{fig:priors.outlier_frac}
\end{figure}

We sought a flexible prior for this work that could reduce to a (nearly) normal
distribution, but had sufficient flexibility to accommodate heavy-tailed
distributions as well. Priors meeting this criterion include:
\begin{itemize}
    \item priors of the form $\nu \sim \Gamma(\alpha, \beta)$ for
          some shape parameter, $\alpha$, and rate parameter, $\beta$ -- e.g.\
          \citet{Juarez:2010} uses $\left\{\alpha = 2, \beta = 0.1\right\}$;
          \citet{Ding:2014} uses $\left\{\alpha = 1, \beta = 0.1\right\}$
    \item A uniform prior in $\frac1\nu \sim U(0, 1)$ \citep{Gelman:2013}
    \item A uniform prior\footnotemark in distribution peak height relative to a
    normal distribution, $t$, such that
    \begin{equation}
        t \equiv
           \sqrt{\frac2\nu}\frac{
                \Gamma\left(\frac{\nu + 1}{2}\right)
            }{
                \Gamma\left(\frac{\nu}{2}\right)
            } \sim U(0, 1).
    \end{equation}
\end{itemize}
These priors, along with the prior we adopted, are illustrated in
Figure~\ref{fig:priors.pdf}. The cumulative distribution functions for the
priors in terms of outlier fraction, $\omega$, are illustrated in
Figure~\ref{fig:priors.outlier_frac}.

\footnotetext{Strictly, \citet{Feeney:2018} approximate this prior with the
closed form prior on shape parameter, $\nu$, of
\begin{equation*}
    \mathcal P\left(\nu\right) \propto
    \frac{
        \Theta(\nu)
    }{
        \left(
            \left(\frac{\nu}{\nu_0}\right)^{1 / (2 a)}
            + \left(\frac{\nu}{\nu_0}\right)^{2 / a}
        \right)^a
    }
\end{equation*}
where $\Theta(\cdot)$ is the Heaviside step function, and $\nu_0$ and $a$ are
constants with values $\sim 0.55$ and $\sim 1.2$ respectively.
}

When testing different priors, we found that priors with significant density in
the range $0 < \nu < 1$ led to sampling difficulties as $\nu$ approached 0;
these low values of $\nu \sim 0.1$ can correspond to outliers that are more than
20 orders of magnitude larger than $\sigma$. In theory, these unphysical regions
of parameter space ought to be excluded during the process of inference.
However, the use of HMC in such cases can lead to divergences in the sampling
process or inefficient sampling as the sampler struggles with regions of high
curvature -- see the discussion of this phenomenon in \citet{Neal:2003} and
\citet{Betancourt:2013}. As discussed in
Section~\ref{sec:formalism.implementation}, reparametrizing the Student's
$t$-distribution as a mixture of normals was found to help with, but not
alleviate, this problem.

Another option would be to place a prior is on outlier fraction $\omega$. It
could be argued that the term outlier loses its meaning when the majority of a
dataset is composed of so-called ``outliers''; therefore, a natural choice of
prior might be a uniform distribution ranging from the normally-distributed
outlier fraction of $\sim$0.00270 to this ``maximum'' outlier fraction of 0.5,
corresponding to $\nu \sim 0.302$. This still leads to large outliers (some of 8
orders of magnitude for 100 draws from the distribution), which are unphysical
and continue to present difficulties when sampling with HMC.

To limit the number of unphysical outliers, we can instead limit the prior to
consider only distributions that are less heavy-tailed than the Cauchy
distribution -- i.e.\ all those with $\nu > 1$. This is the approach taken by
\citet{Gelman:2013}, rendering regions of parameter space inaccessible (in the
case of the \citet{Gelman:2013} prior, $\nu = 1$ is the cutoff, which
corresponds to a Cauchy distribution.) On the other hand, the priors used in
\citet{Juarez:2010, Ding:2014, Feeney:2018} have disproportionate prior density
at low values of $\nu$, which corresponds to models with outliers several orders
of magnitude larger than the predicted effect size.

In this work, we use the prior
\begin{equation}
    \nu \sim \text{Inv-}\Gamma(4, 15),
\end{equation}
where $\text{Inv-}\Gamma(\alpha, \beta)$ is an inverse gamma distribution with
shape parameter, $\alpha$, and scale parameter, $\beta$.

This prior was chosen for two reasons:
\begin{itemize}
    \item the prior is smooth in $\nu$ with no sharp boundaries (unlike
          \citet{Gelman:2013})
    \item the prior has density at a larger range of outlier fractions $\omega$
          than that in \citet{Juarez:2010} but insignificant density
          at unrealistically high outlier fractions, in contrast with the priors
          in \citet{Ding:2014, Feeney:2018}.
\end{itemize}

\section{Estimating variance trade-offs}
\label{sec:fisher}

In Section~\ref{sec:formalism.asymptotic}, we examined the increased standard
deviation in parameter estimates using a toy model, showing the results of
repeated trials alongside the Cram\'er-Rao bound \citep{Rao:1945, Cramer:1946}
in Figure~\ref{fig:formalism.posterior-sd}; here, we derive the formula for this
bound.

For the toy model, we looked at the posterior distribution of the mean $\mu$ for
a series of $N$ points drawn from a normal distribution with zero mean and unit
variance. The log likelihood, $\ell (\mu; \nu, \{x_i\})$, when analyzed with a
$t$-distribution, is
\begin{equation}
    \ell (\mu; \nu, \{x_i\}) =
        - \sum_i \frac{\nu + 1}{2} \log \left(
            1 + \frac{(x_i - \mu)^2}{\nu}
        \right) + C,
\end{equation}
where $C$ is a constant. Thus, the Fisher information is
\begin{equation}
    \mathcal F(\mu; \nu, \{x_i\}) =
        (\nu + 1) \sum_i \frac{\nu - (x_i - \mu)^2}{\nu + (x_i - \mu)^2}.
\end{equation}
We can then evaluate the Cram\'er-Rao bound at the true mean $\mu = 0$,
marginalising over the particular dataset $\{x_i\}$, to derive the variance in
the estimate
\begin{equation}
    {\rm Var}(\hat{\mu}) \geq
        \frac{1}{N(\nu + 1) \left(
            1 - \sqrt{\frac\nu2 \pi} \exp\left(\frac\nu2\right)
            {\rm erfc}\left(\sqrt{\frac\nu2}\right)
        \right)},
\end{equation}
where ${\rm erfc (x)} = 1 - 2/\sqrt{\pi} \int_0^x \exp(-t^2) \,{\rm d}t$ is the
complementary error function.  This bound is used to derive the posterior
standard deviation ratio plotted in Figure~\ref{fig:formalism.posterior-sd}.

\section{Fixed-value calibration data models}
\label{sec:data-models}

Here, we fully specify the dataset models used for fixed-value calibration tests
in Section~\ref{sec:results}.

\subsection{\textit{t}-distributed data}
\label{sec:data-models.t}

In Section~\ref{sec:results.t}, the fixed-value test datasets have $N = 20$
datapoints drawn from the following distribution:
\begin{align}
    x_i \sim&\; \mathcal N (\mu = 2, \sigma^2 = 4) \\
    y_i \sim&\; t_{3} (\mu = 3 + 2 x_i, \sigma^2 = 0.01 / \sigma_{68}^2(\nu = 3)) \\
    \log_{10} \sigma_{x, i} \sim&\; \mathcal N (\mu = -1, \sigma^2 = 0.01) \\
    \log_{10} \sigma_{y, i} \sim&\; \mathcal N (\mu = -0.7, \sigma^2 = 0.01) \\
    \hat{x}_i \sim&\; \mathcal N (\mu = x_i, \sigma^2 = \sigma_{x, i}^2) \\
    \hat{y}_i \sim&\; \mathcal N (\mu = y_i, \sigma^2 = \sigma_{y, i}^2).
\end{align}

\subsection{Normally-distributed data with an outlier}
\label{sec:data-models.outlier}

In Section~\ref{sec:results.outlier}, we generated datasets of $N = 12$ points
using the model:
\begin{alignat}{1}
    x_i& \sim \mathcal N (\mu = 5, \sigma^2 = 9) \\
    y_i& \sim
    \begin{cases}
        \mathcal N (\mu = 3 + 2 x_i - 10, \sigma^2 = 0.04) &
            \text{for the second-largest $x_i$} \\
        \mathcal N (\mu = 3 + 2 x_i, \sigma^2 = 0.04) &
            \text{otherwise} \\
    \end{cases}\\
    \log_{10} \sigma_{x, i}& \sim \mathcal N (\mu = -0.5, \sigma^2 = 0.01) \\
    \log_{10} \sigma_{y, i}& \sim \mathcal N (\mu = -0.3, \sigma^2 = 0.01) \\
    \hat{x}_i& \sim \mathcal N (\mu = x_i, \sigma^2 = \sigma_{x, i}^2) \\
    \hat{y}_i& \sim \mathcal N (\mu = y_i, \sigma^2 = \sigma_{y, i}^2).
\end{alignat}

\subsection{Two-dimensional normal mixture model}
\label{sec:data-models.gmm}

We introduce the parameter $O_i$ to indicate whether the $i$th datapoint is
drawn from the core distribution (in which case, $O_i = 0$) or from the outlier
distribution (for which $O_i = 1$).

In Section~\ref{sec:results.gmm}, we generated a dataset of $N = 200$ points
using the model:
\begin{alignat}{1}
    \boldsymbol{x}_i& \sim
    \begin{cases}
        \mathcal N \left(
            \mu = \begin{pmatrix} -3 \\ 2 \end{pmatrix},
            \Sigma^2 = \begin{pmatrix} 0.5 & -1 \\ -1 & 4 \end{pmatrix}^2
        \right) &
            1 \leqslant i \leqslant 140 \\
        \mathcal N \left(
            \mu = \begin{pmatrix} -1 \\ -1 \end{pmatrix},
            \Sigma^2 = \begin{pmatrix} 1 & 0.2 \\ 0.2 & 0.8 \end{pmatrix}^2
        \right) &
            140 < i \leqslant 200 \\
    \end{cases}\\
    O_i& \sim \mathrm{Bernoulli}(0.1) \\
    y_i& \sim
    \begin{cases}
        \mathcal N (\mu = 2 + (3, 1)^T \cdot x_i, \sigma^2 = 0.16) &
            O_i = 0 \\
        \mathcal N (\mu = 2 + (3, 1)^T \cdot x_i, \sigma^2 = 16.0) &
            O_i = 1 \\
    \end{cases}\\
    \Sigma_{x, i}& \sim \mathcal W_2 (\boldsymbol{V} = 0.1 \mathbb{I}, n = 3) \\
    \log_{10} \sigma_{y, i}& \sim \mathcal N (\mu = -1, \sigma^2 = 0.01) \\
    \hat{x}_i& \sim
        \mathcal N (\mu = \boldsymbol{x}_i, \Sigma^2 = \Sigma_{x, i}^2) \\
    \hat{y}_i& \sim \mathcal N (\mu = y_i, \sigma^2 = \sigma_{y, i}^2),
\end{alignat}
where $\mathcal W_2$ denotes a Wishart distribution over 2x2 matrices and
$\mathbb{I}$ denotes the 2x2 identity matrix.

\subsection{Laplace-distributed data}
\label{sec:data-models.laplace}

In Section~\ref{sec:results.laplace}, we generate $N = 25$ datapoints under the
following model:
\begin{eqnarray}
    x_i &\sim& \mathcal U (-5, 5) \\
    y_i &\sim& \mathrm{Laplace} (\mu = -1 + 0.8 x_i, b = 0.2) \\
    \log_{10} \sigma_{x, i} &\sim& \mathcal N (\mu = -1, \sigma^2 = 0.01) \\
    \log_{10} \sigma_{y, i} &\sim& \mathcal N (\mu = -1, \sigma^2 = 0.01) \\
    \hat{x}_i &\sim& \mathcal N (\mu = x_i, \sigma^2 = \sigma_{x, i}^2) \\
    \hat{y}_i &\sim& \mathcal N (\mu = y_i, \sigma^2 = \sigma_{y, i}^2).
\end{eqnarray}




\bsp	
\label{lastpage}
\end{document}